\documentclass[aps,prb,twocolumn,superscriptaddress]{revtex4-2}

\usepackage{graphicx}
\usepackage{color}
\usepackage[hidelinks]{hyperref}

\usepackage[english]{babel}
\usepackage[utf8]{inputenc}
\usepackage[margin=1in]{geometry}
\usepackage{enumitem}
\usepackage{amsmath,amsfonts}
\usepackage{graphicx}
\usepackage{dcolumn}
\usepackage{bm}
\usepackage{color}
\usepackage{float}
\usepackage{multirow}
\usepackage{tabularx}
\usepackage{mathtools}
\usepackage{wasysym}
\usepackage{latexsym}
\usepackage{hhline}
\usepackage{changepage}

\begin{document}

\title{Understanding the Anomalous Hall effect in Co$_{1/3}$NbS$_{2}$ from crystal and magnetic structures}



\author{K. Lu}
\affiliation{Department of Physics, University of Illinois at Urbana-Champaign, Urbana, IL 61801, USA}
\affiliation{Materials Research Laboratory, University of Illinois at Urbana-Champaign, Urbana, IL 61801, USA}
\author{A. Murzabekova}
\affiliation{Department of Physics, University of Illinois at Urbana-Champaign, Urbana, IL 61801, USA}
\affiliation{Materials Research Laboratory, University of Illinois at Urbana-Champaign, Urbana, IL 61801, USA}
\author{S. Shim}
\affiliation{Department of Physics, University of Illinois at Urbana-Champaign, Urbana, IL 61801, USA}
\affiliation{Materials Research Laboratory, University of Illinois at Urbana-Champaign, Urbana, IL 61801, USA}
\author{J. Park}
\affiliation{Department of Materials Science and Engineering, University of Illinois at Urbana-Champaign, Urbana, IL 61801, USA}
\affiliation{Materials Research Laboratory, University of Illinois at Urbana-Champaign, Urbana, IL 61801, USA}
\author{S. Kim}
\affiliation{Department of Physics, University of Illinois at Urbana-Champaign, Urbana, IL 61801, USA}
\affiliation{Materials Research Laboratory, University of Illinois at Urbana-Champaign, Urbana, IL 61801, USA}
\author{L. Kish}
\affiliation{Department of Physics, University of Illinois at Urbana-Champaign, Urbana, IL 61801, USA}
\affiliation{Materials Research Laboratory, University of Illinois at Urbana-Champaign, Urbana, IL 61801, USA}
\author{Y. Wu}
\affiliation{Neutron Scattering Division, Oak Ridge National Laboratory, Oak Ridge, Tennessee 37831, USA}
\author{L. DeBeer-Schmitt}
\affiliation{Neutron Scattering Division, Oak Ridge National Laboratory, Oak Ridge, Tennessee 37831, USA}
\author{A. A. Aczel}
\affiliation{Neutron Scattering Division, Oak Ridge National Laboratory, Oak Ridge, Tennessee 37831, USA}
\author{A. Schleife}
\affiliation{Department of Materials Science and Engineering, University of Illinois at Urbana-Champaign, Urbana, IL 61801, USA}
\affiliation{Materials Research Laboratory, University of Illinois at Urbana-Champaign, Urbana, IL 61801, USA}
\affiliation{National Center for Supercomputing Applications, University of Illinois at Urbana-Champaign, Urbana, Illinois 61801, USA}
\author{N. Mason}
\affiliation{Department of Physics, University of Illinois at Urbana-Champaign, Urbana, IL 61801, USA}
\affiliation{Materials Research Laboratory, University of Illinois at Urbana-Champaign, Urbana, IL 61801, USA}
\author{F. Mahmood}
\affiliation{Department of Physics, University of Illinois at Urbana-Champaign, Urbana, IL 61801, USA}
\affiliation{Materials Research Laboratory, University of Illinois at Urbana-Champaign, Urbana, IL 61801, USA}
\author{G. J. MacDougall}
\email{gmacdoug@illinois.edu}
\affiliation{Department of Physics, University of Illinois at Urbana-Champaign, Urbana, IL 61801, USA}
\affiliation{Materials Research Laboratory, University of Illinois at Urbana-Champaign, Urbana, IL 61801, USA}

\date{\today}

\begin{abstract}
A large anomalous Hall effect (AHE) has recently been observed in the intercalated transition metal dichalcogenide (TMDC) Co$_{1/3}$NbS$_{2}$ below a known magnetic phase transition at $T_N$ = 29 K. The spins in this material are widely believed to order in a highly symmetric collinear antiferromagnetic configuration, causing extensive debate about how reports of an AHE can be reconciled with such a state. In this article, we address this controversy by presenting new neutron diffraction data on single crystals of Co$_{1/3}$NbS$_{2}$ and an analysis that implies that moments in this material order into a non-collinear configuration, but one that maintains the same refelction symmetries as the collinear phase. We present new transport and magneto-optic Kerr measurements which show that AHE signatures persist below $T_N$ to temperatures as low as $T$ = 5 K and firmly associate them with the long-range antiferromagnetic order. Finally, we show that these AHE signatures can be quantitatively reproduced by density functional theory (DFT) calculations based on the lattice and spin state determined with neutron diffraction. These combined findings establishes the veracity of the 'crystal Hall effect' picture, which shows how such effects can emerge from the shape of magnetic orbitals in compounds containing chiral lattice symmetry regardless of the symmetry of the ordered spin configuration. These results illuminate a new path for the discovery of anomalous Hall materials and motivate a targeted study of the transport properties of intercalated TMDCs and other compounds containing antiferromagnetic order and chiral lattice symmetry.

\end{abstract}

\maketitle

The anomalous Hall effect (AHE) is an intrinsic transport property which is of central importance for modern studies of topology in magnetic materials\cite{nagaosa_2010_anomalous}. Characterized by a transverse conductance in the absence of an applied magnetic field, it has traditionally been considered a property of ordered ferromagnets \cite{mathieu2004scaling}. Recent developments however have used the concept of Berry curvature to more generally interpret an intrinsic AHE as a momentum-space geometrical effect, similar to the quantum Hall effect and the transport properties of topological insulators \cite{sundaram_1999_wavepacket}. Consequently, there has been a surge of investigations into the Hall physics of more complex magnetic compounds, including frustrated \cite{kalitsov_2009_anomalous, udagawa_2013_anomalous}, chiral \cite{ohgushi_2000_spin, a2021_chiralitydriven, taillefumier_2006_anomalous, lee_2007_hidden}, noncollinear \cite{takatsu_2010_unconventional}, and noncoplanar antiferromagnetic systems \cite{shindou_2001_orbital}. Examples from materials with zero-magnetization spin states are particularly interesting, as they are often unexpected and invariably expand our understanding of the origin and significance of AHE signatures.

\begin{figure}[t]
\includegraphics[width=0.5\textwidth]{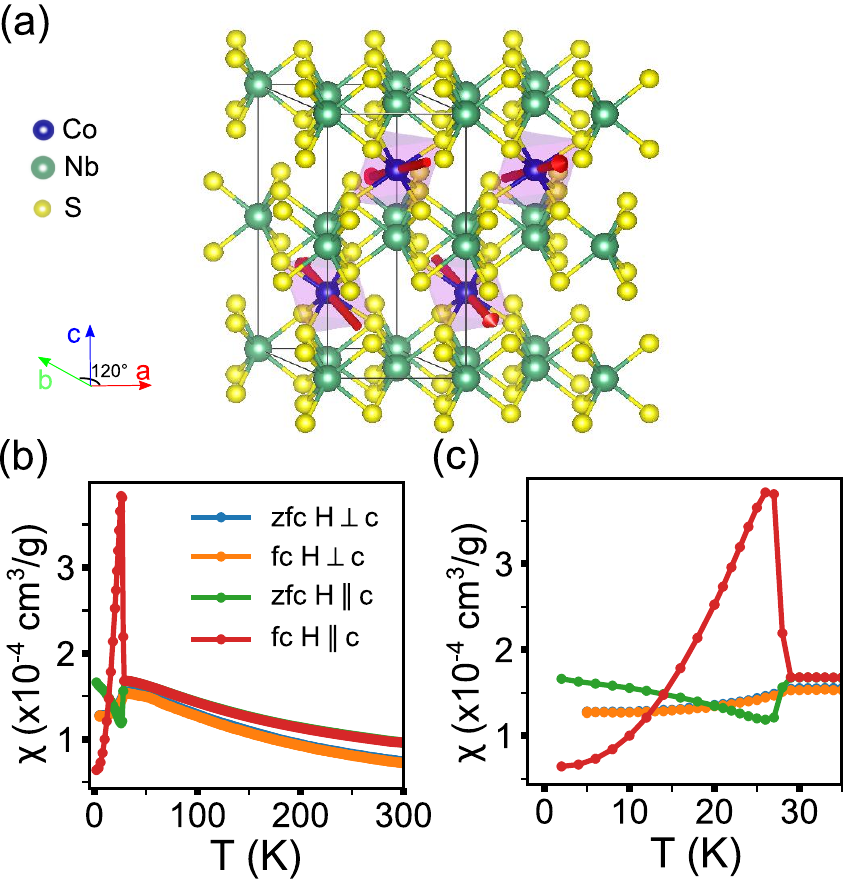}
 \caption{\label{fig:mag_struc} (a) The magnetic unit cell of Co$_{1/3}$NbS$_{2}$ based on our single crystal neutron diffraction data. Solid black lines represent the chemical unit cell, and arrows represent the fitted spin structure. (b) The zfc and fc magnetic susceptibility in both in-plane and out-of-plane directions. Applied fields were H = 500 Oe and 100 Oe, respectively. (c) The susceptibility in the low temperature region. }
\end{figure}

A current high-profile example is the unexpectedly large AHE recently observed in the intercalated transition metal dichalcogenide (I-TMDC) Co$_{1/3}$NbS$_{2}$ below the magnetic ordering transition at $T_N$ = 29K \cite{ghimire_2018_large, tenasini_2020_giant}. This is an antiferromagnetic material from the I-TMDC family T$_{1/3}$MS$_{2}$, where magnetic $T$ cations are intercalated between quasi-2D $M$S$_2$ planes and occupy the 2c Wyckoff site of the non-centrosymmetric space group, P6$_{3}$22 \cite{ghimire_2013_magnetic, ghimire_2018_large, nair_2019_electrical, karna_2019_consequences, lu2020canted}, as shown in Fig.~\ref{fig:mag_struc}(a). Several compounds in this family have recently attracted great attention for new and unusual physics, including the discovery of exotic chiral spin textures \cite{ghimire_2013_magnetic}, magnetic switching behavior \cite{nair_2019_electrical}, and metallic antiferromagnetism with the potential for functional applications \cite{siddiqui_2020_metallic}. In Co$_{1/3}$NbS$_{2}$, reports of AHE have generated considerable excitement, largely because long-standing neutron diffraction measurements have identified the magnetism in this material with a highly symmetric collinear antiferromagnetic phase \cite{parkin_1983_magnetic}. Since the associated magnetic space group (MSG) shows $\mathcal{C}_{2}$ rotational symmetries in all three lattice directions, a traditional analysis predicts an identically zero AHE in this state \cite{suzuki2017cluster}.

This apparent inconsistency has inspired several unconventional theoretical explanations and motivated a careful reevaluation of known experimental data. Among the ideas being discussed are a new multi-component non-collinear spin structure which breaks centrosymmetry \cite{tenasini_2020_giant, heinonen2022magnetic, yanagi2022generation} or possible 2D electronic bands with non-trivial topology\cite{tenasini_2020_giant}. Multiple groups have suggested a role for the weak ferromagnetic moment reported in this material directly below the ordering transition \cite{ghimire_2018_large}, perhaps enhanced by topological Weyl points in the band structure \cite{chang2018topological,tanaka2022large}. At least one study has noted that collinear antiferromagnetism can lead to AHE in a chiral crystal structure due to the effect of non-magnetic atoms on the shape of the magnetic orbitals- an idea dubbed the 'crystal Hall effect'- but that study was unable to reproduce the size of the effect in Co$_{1/3}$NbS$_{2}$ without hypothesizing a significant number of sulfur vacancies\cite{smejkal_2019_crystal}. Additional data is needed settle this debate.

To address this controversy, we have grown single crystals of Co$_{1/3}$NbS$_{2}$ and studied them with neutron diffraction, magneto-transport, and magneto-optic Kerr spectroscopy measurements. Our transport and optical measurements confirm the existence of a large AHE in our crystals of comparable magnitude to previous results, but extended to temperatures well below the limited region where ferromagnetism is observed. Our neutron diffraction results are interpreted using a modern symmetry analysis and unambiguously imply a non-collinear ordered spin state. In particular, we show that our neutron data are best described by a single-\textbf{k} magnetic structure with spins in the chemical unit cell containing parallel components along the b-axis and an anti-parallel tilting out-of-plane (Fig.~\ref{fig:mag_struc}). The MSG for this structure is P$_C$2$_1$2$_1$2$_1$, which again has $\mathcal{C}_{2}$ symmetries in all three lattice directions. Nevertheless, we show that DFT calculations using the output from our neutron refinement predicts a Kerr ellipticity which matches our independent measurements. This result implies that the AHE in stoichiometric crystals of Co$_{1/3}$NbS$_{2}$ can be understood entirely within the crystal Hall effect picture if the correct magnetic structure is taken into account.

Single crystal samples of Co$_{1/3}$NbS$_{2}$ were grown at the University of Illinois using the standard 2-stage chemical vapor transport (CVT) technique reported in Ref.~\cite{ghimire_2013_magnetic} and characterized using equipment in the Illinois Materials Research Laboratory. The sample composition was verified by the Energy Dispersive Spectroscopy (EDS) mode of the Joel 7000 Scanning Electron Microscope and confirmed an ideal ratio of Co:Nb:S = 0.33:1:2 for several different samples from the growth.

The magnetic properties were first characterized using a Quantum Design MPMS3 magnetometer. Magnetic susceptibility was measured as a function of temperature, with crystals mounted both with the c-axis parallel and perpendicular to the applied field direction. The resulting zero-field-cooled (zfc) and field-cooled (fc) magnetic susceptibilities are shown in Fig.~\ref{fig:mag_struc}(b) and (c) and are similar to what has been reported previously \cite{ghimire_2018_large}. The N\'eel temperature is $T_{N}$ = 29 K, as determined by the abrupt downturn in the susceptibility. In the out-of-plane direction, there is a sharp increase in the fc magnetic susceptibility immediately below T$_{N}$, in line with the 0.001 $\mu_B$ ferromagnetic moment seen by others \cite{ghimire_2018_large}. This moment quickly disappears below $T$ = 25 K, and the ground state is consistent with a zero moment antiferromagnetic state. The susceptibility at high temperatures follows a Curie-Weiss temperature dependence. Fitting the fc H$\parallel$c data in the region T $>$ 200 K to $\chi = p_{\mathrm{eff}}^2/3(T-\Theta_{W})+B$ gave a Weiss temperature $\Theta_{W} = -160(5)$ K and an effective moment $p_{\mathrm{eff}}$ = 3.16 $\pm$ 0.05 $\mu_{B}$/Co, consistent with a Co$^{2+}$ valence state.

\begin{figure}[t]
\includegraphics[width=0.5\textwidth]{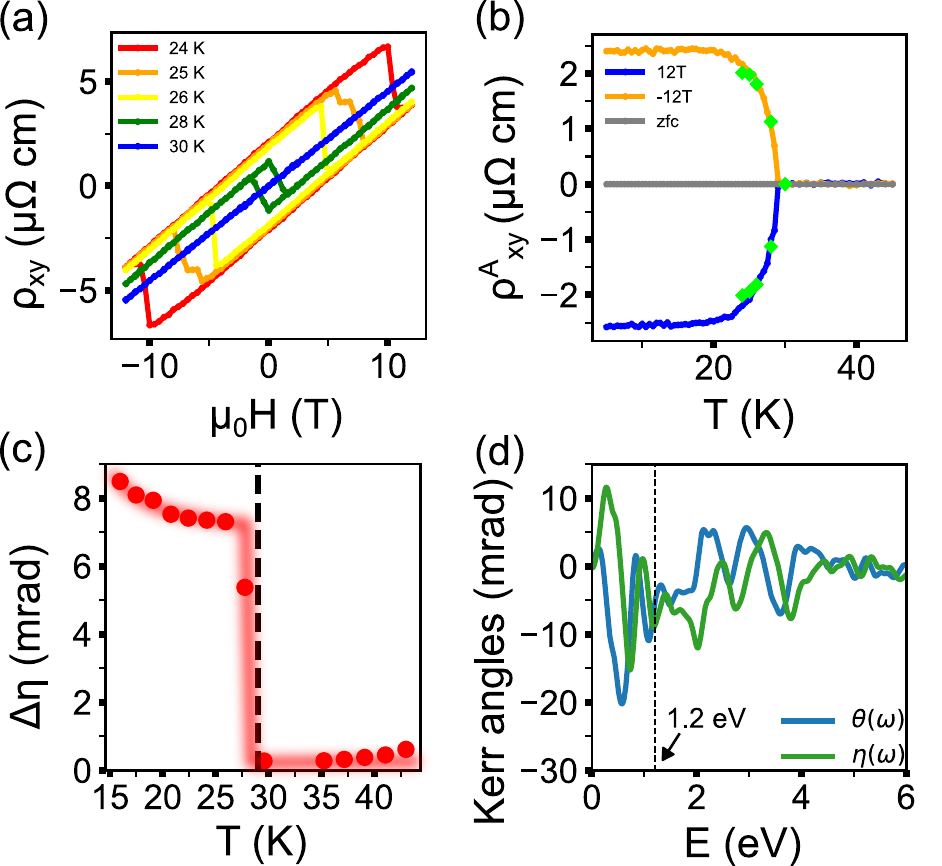}
\caption{\label{fig:AHE_Moke} (a) The transverse resistivity $\rho_{xy}$ vs the applied field from our exfoliated crystals, and (b) a plot of $\rho^{A}_{xy}$ vs temperature obtained from zfc-fc temperature sweeps. The lime-colored diamond markers are data points extracted from the field-sweeps in panel (a). (c) The change in Kerr ellipticity as a function of temperature (red dots). The red shading is a guide to the eye. (d) The DFT determination of Kerr angles (real part $\theta$ and imaginary part $\eta$) as a function of photon energy using lattice and magnetic structures from the current work. The dashed line indicates the energy ($\sim$1.2 eV) used in our MOKE measurements. }
\end{figure}

Electronic transport properties were studied using a fabricated device based on an exfoliated Co$_{1/3}$NbS$_{2}$ crystal (thickness t = 44 nm, lateral dimension 1.6$\times$4 $\mu$m$^{2}$). In-plane transverse resistivity $\rho_{xy}$ was measured using an Oxford TeslatronPT Cryofree superconducting magnet system, and the AHE component $\rho^{A}_{xy}$ was extracted using two different methodologies\cite{ghimire_2018_large, tenasini_2020_giant, nakatsuji2015large, kiyohara2016giant}. In the first, we measured transverse resistivity $\rho_{xy}$ while sweeping the magnetic field, $H\parallel$c in the range -12 T to +12 T with the results plotted in Fig.~\ref{fig:AHE_Moke}(a). For temperatures $T > T_{N}$, $\rho_{xy}$ shows a $H$-linear ordinary Hall effect (blue curve). As $T$ is lowered below the transition (e.g. green curve measured at $T$ = 28 K), a pronounced hysteresis is manifested, such that $\rho_{xy} \neq 0$ at $H$ = 0. The AHE component $\rho^{A}_{xy}$ was then obtained by subtracting the $H$-linear component determined at $T$ = 30 K from $\rho_{xy}$. When lowering $T$, the coercive magnetic field $H_{c}$, rapidly increases and soon becomes larger than the highest field accessible in our experimental setup (12 T). The results in Figs.~\ref{fig:AHE_Moke}(a) are similar to those presented in Ref.~\cite{ghimire_2018_large}, but we are able to confirm AHE behavior to temperatures as low as $T$ = 24 K due to the larger field range of the current instrument.

To extend the temperature range further, we followed with zfc-fc temperature sweeps from well above $T_{N}$ down to 5 K using a field of $H$ = 12 T. $\rho^{A}_{xy}$ was inferred by subtracting the zfc component from the fc transverse resistivity $\rho_{xy}$ \cite{tenasini_2020_giant, nakatsuji2015large, kiyohara2016giant}. As shown in Fig.~\ref{fig:AHE_Moke}(b), the AHE extracted this way is assumed to be zero during zfc (grey) due to balanced contributions from different antiferromagnetic domains, while fc curves (blue and orange) reveal a large finite AHE component at all temperatures below $T_{N}$. The AHE component read from field sweeps in Fig.~\ref{fig:AHE_Moke}(a) are included in Fig.~\ref{fig:AHE_Moke}(b) as the lime diamond markers, demonstrating consistency between the two methodologies in the relevant temperature range. The anomalous Hall conductivity $\sigma^{A}_{xy} = \frac{\rho^{A}_{xy}}{\rho^{2}_{xx}+\rho^{2}_{xy}}$ in our devices reached up to 16.8 S/cm at 5 K, comparable to the maximum value 27 S/cm measured in Co$_{1/3}$NbS$_{2}$ bulk crystals \cite{ghimire_2018_large}. Importantly, our data also allow us to unambiguously associate the AHE in this material with the antiferromagnetic order at base temperature, and not the weak ferromagnetism closer to the transition.

We supplemented these results with magneto-optic Kerr effect (MOKE) measurements, which can be thought of as an optical analogue to transverse conductivity measurements (see Supplemental Materials). A standard tool to characterize order in ferromagnets \cite{gong2017discovery}, MOKE is increasingly used to extract important information about various antiferromagnetic spin systems \cite{feng_2015_large, siddiqui_2020_metallic, higo_2018_large, little_observation, kang2022}. Of particular relevance, it has been shown that polar MOKE is expected to provide prominent signatures in collinear antiferromagnetic metals with a chiral crystal structure due to the non-zero Berry curvature of occupied electronic bands \cite{zhou_2021_crystal}. In our measurements, a linearly polarized laser beam (1030 nm, 76 MHz, 50 mW) was focused on the sample surface with a spot size of 70 $\mu$m (1/e$^2$ width), and the reflected beam was picked up and modulated by a photo-elastic modulator (Hinds PEM-100) operating at 50 kHz. This allowed a measurement of the imaginary part (i.e. Kerr ellipticity) of the Kerr angle ($\Psi = \theta + i\eta$) with a precision of $<$10 $\mu$rad. Fig.~\ref{fig:AHE_Moke}(c) presents the resulting ellipticity values as a function of temperature and, much like transport results, shows a sudden jump at $T = T_N$ from zero to a large Kerr angle more typical for ferromagnets.

\begin{figure}[h]
\includegraphics[width=0.5\textwidth]{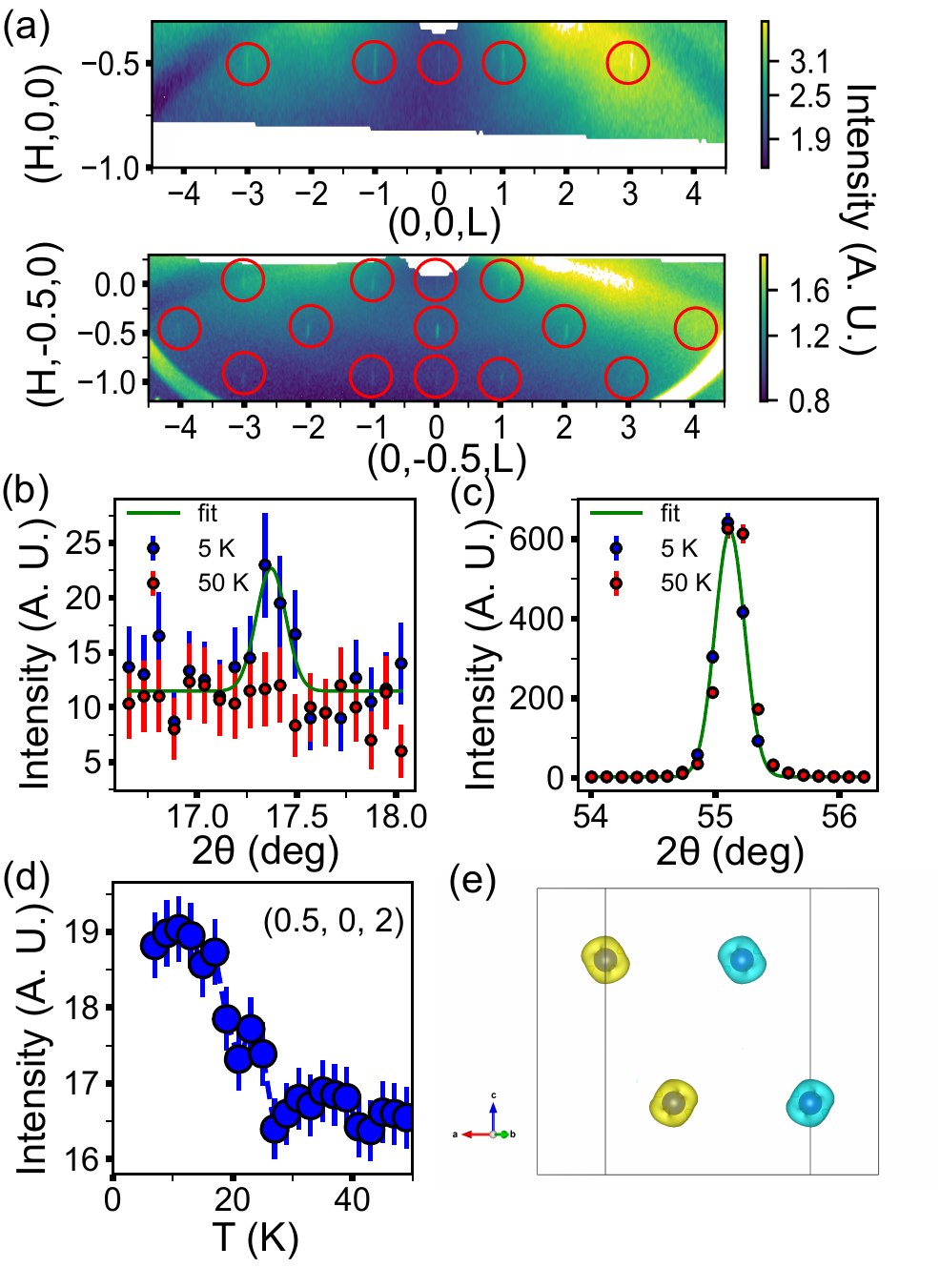}
\caption{\label{fig:fit} (a) Plots of raw neutron scattering intensity in the (H, 0, L) and (H, -0.5, L) reciprocal lattice planes measured using WAND$^{2}$. The weak peaks seen are all magnetic. Panels (b) and (c) show rocking curves of the magnetic Bragg peak (0.5, 0, 2) and nuclear Bragg peak (3, 0, 0) at 5 K (blue) and 50 K (red), respectively, measured using DEMAND. (d) A plot of scattering intensity of the magnetic Bragg peak (0.5, 0, 2) vs temperature. (e) A plot of the magnetization density in a single magnetic unit cell, predicted by DFT calculations described in the main text.}
\end{figure}

Our neutron diffraction measurements were performed on a 10 mg single crystal using both the DEMAND and WAND$^{2}$ instruments at the High Flux Isotope Reactor of Oak Ridge National Laboratory. For both experiments, the sample was glued to an aluminum holder and mounted in either a closed cycle refrigerator (DEMAND) or a cryostat (WAND$^{2}$) with base temperatures of 4 K and 1.5 K, respectively. For the DEMAND experiment, a monochromatic incident beam of wavelength 1.546 \AA\ was selected with a bent Si-220 monochromator. The nuclear and magnetic intensities were captured by rocking curve scans around each peak position. For the WAND$^{2}$ experiment, we mounted the sample in the (H,H,L) scattering plane. A neutron wavelength of 1.486 \AA\ was selected using a Ge monochromator. A large region of reciprocal space near the (H, H, L) plane with 3$< 2\theta <$ 260$^\circ$ was mapped out through a full sample rotation. Fig.~\ref{fig:fit}(a) shows a subset of these data plotted in the (H, 0, L) and (H, -0.5, L) scattering planes.

Both DEMAND and WAND$^{2}$ experiments confirmed the expected hexagonal lattice structure with comparable values for a = b = 5.741(3) \AA\ and c = 11.866(5) \AA\ (1.5 K). Both experiments also confirmed a $\mathbf{k} = (0.5, 0, 0)$ propagation vector for the magnetic structure, as seen previously\cite{parkin_1983_magnetic} and discussed extensively below. Three magnetic domains are allowed by symmetry for this propagation vector, $(0.5, 0, 0)$, $(0, -0.5, 0)$ and $(-0.5, 0.5, 0)$, and their signals were fitted simultaneously during analysis. For the DEMAND data, nuclear Bragg intensities were extracted at all temperatures by fitting the peaks to a Gaussian profile with constant background. For half-integer positions, magnetic scattering intensities were derived from the difference between low and high temperature signals, to account for contamination from $\lambda/2$ scattering from nuclear peaks at integer positions. The WAND$^2$ data has no higher order wavelength contamination, and an ellipsoidal integration technique was used to extract both nuclear and magnetic Bragg peak intensities from 3D reciprocal space data after normalizing to a vanadium standard.

In Fig.~\ref{fig:fit}(b) and (c), we show the line cuts of scattering intensities at $(0.5, 0, 2)$ and $(3, 0, 0)$, respectively, integrated over tilt directions at temperatures $T$ = 5 K and 50 K. There is an increase in scattering with decreasing temperature at the half-integer position but none at the integer position, consistent with the development of magnetic order with propagation vector $\mathbf{k} = (0.5, 0, 0)$. The scattering at $(-0.5, 0, L)$ positions in the WAND$^2$ data in Fig.~\ref{fig:fit}(a) reaffirms this conclusion, while intensity at $(H, -0.5, L)$ positions reveals comparable scattering from the other magnetic domains. Fig.~\ref{fig:fit}(d) shows the extracted intensity of the magnetic $(0.5, 0, 2)$ peak as a function of temperature and confirms the known ordering transition at $T_{N}$ = 29 K. Though allowed by symmetry, we saw no sign of incommensuration of any Bragg peaks, ruling out a spin spiral state with a wavelength shorter than $\sim$450 \AA\ .

Refinements of the neutron diffraction patterns were performed using the Jana2020 software \cite{petek_2014_crystallographic}. The nuclear structure factors from each instruments were fitted independently, assuming a P6$_{3}$22 structural model with Co at the 2c Wyckoff sites. The fits only varied the overall scale factors, atomic positions, and isotropic atomic displacement parameters. With this model, we were able to describe the nuclear structure factors F$^{2}_{nuclear}$ with an R-factor of 5.8\% (DEMAND) and 4.6\% (WAND$^2$). We demonstrate the quality of the fit by plotting the observed structural factors against the calculated ones in the Supplementary Materials. The space group, lattice parameters, and sulfur positions are all consistent with literature values \cite{vanlaar_1971_magnetic}.

Suitable MSGs were generated using the ISODISTORT software \cite{isodistort_1, isodistort_2}. Refinement of magnetic scattering intensity was performed assuming each of the four MSGs in turn, with a fixed scale factor, atomic coordinates based on the nuclear structural refinements, and three equally-populated magnetic domains. Of the four models consistent with a single propagation vector $\bf{k}$ (see Supplemental Materials), only the non-collinear magnetic model P$_C$2$_{1}$2$_{1}$2$_{1}$ is able to describe the general pattern of magnetic scattering intensity with any degree of accuracy. This model is equivalent to the $\Gamma_{2}$ irreducible representation (IR) of the little group G$_{\bf{k}}$ and is characterized by nearly antiparallel spins restricted to the bc-plane with neighboring spins alternately tilting above and below the ab-plane (see Fig.~\ref{fig:mag_struc}). Fits assuming other MSGs gave goodness-of-fit parameters, R $>$60\% and moreover predicted zero scattering intensity at multiple locations where peaks were observed in our data. Qualitatively, the correctness of the $\Gamma_{2}$ model is also suggested by the strong intensity observed at the (-0.5, -0.5, 0) position and non-zero intensity observed at $(0, -0.5, 0)$, as noted elsewhere \cite{wu2022highly}. The ordered moment from this model is refined to be (0, 1.929(1), 0.35(1)) $\mu_{B}$ in hexagonal coordinates, corresponding to moment size 1.96(6)$\mu_B$ and a tilt angle of $\sim$10$^\circ$ from the ab-plane. More sophisticated magnetic models with multiple $\bf{k}$-vectors were seen to result in poorer quality fits with unphysical moment sizes.

These conclusions about the magnetic ordering constitute a major deviation from the results of previous neutron diffraction work. Specifically, the classic work of Parkin \textit{et al.} \cite{parkin_1983_magnetic} concluded a collinear ordered state akin to the IR $\Gamma_{4}$, a model which we have rejected with great discrimination. The reason for this discrepancy is unclear, but it may be related to small differences in the choice of measured Bragg peaks, refined nuclear structure, and assumptions about the form factor for Co$^{2+}$ \cite{parkin_1983_magnetic}. Our confidence in our model however is based on qualitative and quantitative considerations of our two independent data sets and is only reinforced by our density functional theory (DFT) calculations, described below.

To correlate our scattering and AHE results, we performed first-principles DFT \cite{DFT, Kohn-Sham} simulations using the Vienna \textit{Ab-Initio} Simulation Package (VASP) \cite{vasp1, vasp2}. The exchange and correlation interactions were described using the generalized-gradient approximation, as parameterized by Perdew, Burke, and Ernzerhof \cite{PBE}. Kohn-Sham states were expanded into a plane-wave basis up to a kinetic-energy cutoff of 600 eV. A 5\,$\times$\,10\,$\times$\,5 Monkhorst-Pack \textbf{k}-point grid \cite{MP} was used for sampling of the Brillouin zone. Using these parameters converges the Kerr rotation to an accuracy better than 4.8 mrad and ellipticity better than 4.3 mrad up to photon energies of 10 eV, where the maximum magnitude of the Kerr rotation and ellipticity is 20.6 mrad and 15.3 mrad, respectively.

The magnetic configuration of the material was described using a non-collinear framework, including spin-orbit coupling \cite{vasp-mag}. After initializing the magnetic unit cell in MSG P$_C$2$_1$2$_1$2$_1$, as determined by neutron diffraction, we relaxed the magnetic moments within DFT and obtained a similar final state with an order moment smaller by 0.52 $\mu_B$ and a marginally larger tilt of $14.6^{\circ}$.  We then computed the complex, frequency-dependent dielectric tensor \cite{vasp-opt}, where Drude like intraband contributions in the low-energy range of the dielectric tensors are not included. Using the diagonal and off-diagonal components of the dielectric tensor, we computed the complex polar Kerr spectra following Ref.\ \cite{pmoke}. Further details and full results are given in the Supplemental Materials.

The MOKE parameters predicted by DFT are shown in Fig.~\ref{fig:AHE_Moke}(d) and imply a Kerr ellipticity of $-9$ mrad at $E$ = 1.2eV, very close to the measured value of -$8.7$ mrad shown in Fig.~\ref{fig:AHE_Moke}(c). This is a significant result, since MOKE is an optical analogue of the transport AHE which is directly related to the symmetry of the bands and less susceptible to sample-specific electron mobility effects. The P$_C$2$_{1}$2$_{1}$2$_{1}$ MSG determined from our neutron data should forbid an AHE by symmetry if one focusses on spins alone. Therefore, the success of our DFT calculations should be attributed to the crystal Hall effect, where pertinent global symmetries are broken not by the ordered spins but by the shape of the magnetic orbitals, which are in turn determined by the placement of the non-magnetic ligand atoms in the chiral crystal structure. This picture is reinforced by the plot of magnetization density shown in Fig.~\ref{fig:fit}(e), where one can see that the combination of time-reversal and reflection symmetry is preserved by the spins but not the orbital about the cobalt positions.

To summarize then, our collective results paint a self-consistent picture that explains the appearance of a large AHE in Co$_{1/3}$NbS$_{2}$. We have presented transport and optical evidence that unambiguously associate the AHE with the low-temperature antiferromagnetic state in this material, and not the weak ferromagnetic moment that appears only near the transition temperature. The analysis of our neutron diffraction data reveals that spins in this state order in a non-collinear configuration described by the magnetic space group P$_C$2$_1$2$_1$2$_1$. Moreover, the success of DFT in explaining our MOKE data using experimentally derived spin and atomic positions favors a crystal Hall origin for the AHE in this material, even in stoichiometric samples. The success of the crystal Hall picture in explaining exotic electronic and optical phenomena in Co$_{1/3}$NbS$_{2}$ should motivate researchers to look beyond spin symmetries and look for AHE signatures in a variety of other magnetic materials with chiral lattice structures.

\begin{acknowledgments}
This work was sponsored by the National Science Foundation under Grant No. DMR-1455264-CAR (K.L., L.K. and G.J.M) and through the University of Illinois Materials Research Science and Engineering Center DMR-1720633 (A.M., S.S., J.P., S.K., A.S., N.M. and F.M). K.L. would particularly like to thank V\'{a}clav Pet\v{r}\'{i}\v{c}ek (petricek@fzu.cz) for his dedication to developing and customizing multiple-$\bf{k}$ refinement in Jana2020. Synthesis, fabrication, transport, and magnetization measurements were carried out in part in the Materials Research Laboratory Central Research Facilities, University of Illinois. Computational work made use of the Illinois Campus Cluster, a computing resource that is operated by the Illinois Campus Cluster Program (ICCP) in conjunction with the National Center for Supercomputing Applications (NCSA) and which is supported by funds from the University of Illinois at Urbana–Champaign. Neutron scattering work used resources at the High Flux Isotope Reactor, a DOE Office of Science User Facility operated by the Oak Ridge National laboratory.
\end{acknowledgments}


%

\break

\frenchspacing

\begingroup
    \fontsize{24pt}{24}\selectfont
    \begin{verbatim}
        SUPPLEMENTAL MATERIALS
    \end{verbatim}
\endgroup

\section{Linear Magneto-optic Kerr effect (MOKE)}
The magneto-optical Kerr effect (MOKE) is a powerful tool to probe crystal magnetization. When time reversal symmetry in a material is broken due to the presence of an external magnetic field or intrinsic magnetization, linearly polarized light becomes elliptically polarized upon reflection from the sample surface. While MOKE is a standard tool to characterize magnetic ordering in ferromagnets, it can also be used to detect noncollinear magnetic structure in antiferromagnets \cite{siddiqui2020metallic, higo2018large}. Additionally, it has been shown that MOKE can also occur in collinear antiferromagnetic metals with a chiral crystal structure due to a non-zero Berry curvature of the occupied electronic bands \cite{zhou2021crystal}.

In this work, we perform polar MOKE in which the light is normally incident upon the sample a-b plane. As the lattice of Co$_{1/3}$NbS$_2$ corresponds to space group $P6_322$ [182], the dielectric tensor can be written in terms of the diagonal components $\epsilon_{xx}$ and $\epsilon_{zz}$, and the off-diagonal components $\epsilon_{xy}$ as follows:

\begin{equation*}
\epsilon =
   \begin{pmatrix}
    \epsilon_{xx} & \epsilon_{xy} & 0 \\
    -\epsilon_{xy} & \epsilon_{xx} & 0 \\
    0 & 0 & \epsilon_{zz} \\
    \end{pmatrix}
\end{equation*}
Here the coordinate axes have been taken to lie along the principal axes of the crystal. A complete description of MOKE is then given by the complex refractive index:
\begin{equation*}
N(\omega) = \sqrt{\epsilon(\omega)} = n(\omega) + ik(\omega)
\end{equation*}
for light with a given polarization propagating along certain directions in the sample. Here $n$ and $k$ are the refractive index and the extinction coefficient, respectively. For circularly polarized light that is normally incident on the sample (propagating along the z-axis), the complex refractive index is given by:

\begin{equation*}
N_{\pm} = n_{\pm} + ik_{\pm} = \sqrt{\epsilon_{xx} \pm \epsilon_{xy}}
\end{equation*}
here $\pm$ denotes left and right circularly polarized light. Note that linearly polarized light can always be written as a combination of right and left circularly polarized light and so the polar complex Kerr angle ($\theta_K + i\eta_K$) for normally incident linearly polarized light can be determined from:

\begin{equation*}
\frac{1+\mathrm{tan} \eta_K}{1-\mathrm{tan}\eta_K} e^{2i\theta_K} = \frac{(1+N_{+})(1-N_{-})}{(1-N_{+})(1+N_{-})}
\end{equation*}
For small $\theta_K$ and $\eta_K$, the expression above can be approximated as:
\begin{equation*}
\theta_K + i\eta_K \sim \frac{-\epsilon_{xy}}{(\epsilon_{xx} - 1)\sqrt{\epsilon_{xx}}}
\end{equation*}

The optical conductivity tensor $\sigma_{ij}$ is related to the dielectric tensor $\epsilon_{ij}$ as:

\begin{equation*}
\epsilon_{ij}(\omega) = \delta_{ij} + \frac{4 \pi i}{\omega}\sigma_{ij}(\omega)
\end{equation*}

Thus, the complex Kerr angle is directly related to the off-diagonal components of the conductivity tensor (i.e., the Hall conductivity) as follows:
\begin{equation*}
\theta_K + i\eta_K \sim \frac{-\sigma_{xy}}{\sigma_{xx}\sqrt{1+i(4\pi/\omega)\sigma_{xx}}}
\end{equation*}
Note that the above is a general result that applies whenever light is normally incident onto a sample surface with greater than three-fold rotational symmetry \cite{Kahn_1969}.

\section{Irreducible Representations for  $\mathbf{k} = (0.5, 0, 0)$}

\begin{figure*}[h]
\includegraphics[width=0.75\textwidth]{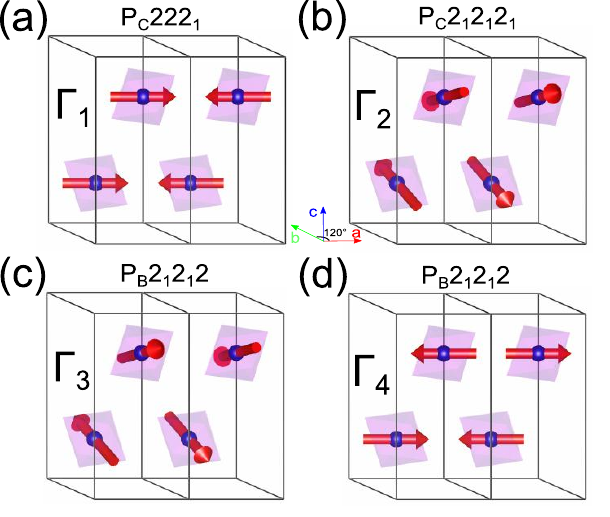}
\caption{\label{fig:state} Plots of the four irreducible representations (IRs) associated with the P6$_{3}$22 spacegroup and a $\mathbf{k} = (0.5, 0, 0)$ propagation vector. Arrows denote Co$^{2+}$ spin directions, and octahedra denote the local crystal environments set by sulfur positions. All models are displayed in the hexagonal chemical unit cell and show antiparallel spins between neighboring chemical unit cells in the a-axis direction. $\Gamma_{1}$ and $\Gamma_{4}$ are two different in-plane antiferromagnetic models with moments parallel to the a-axis. $\Gamma_{2}$ and $\Gamma_{3}$ are canted models, where spins lie in the bc-plane. The canting angle away from the ab-plane is a free parameter in these IRs which is fit during refinement. The MSG for each IR is labelled at the top the respective panel. Plots were made using the VESTA structure visualization program \cite{momma2011vesta}. }

\end{figure*}

The magnetically ordered state in Co$_{1/3}$NbS$_{2}$ is described by the propagation vector $\bf{k}$ = (0.5,0,0). Assuming moments on cobalt sites of the lattice structure described in the main text, the magnetic space groups associated with this value of $\bf{k}$ were generated systematically using the ISODISTORT software\cite{isodistort_1, isodistort_2}. The simplest models with only one propagation vector, $\bf{k}$, were examined closely. The four different magnetic space groups consistent with the single propagation vector, $\bf{k}$, are P$_C$222$_{1}$, P$_C$2$_{1}$2$_{1}$2$_{1}$, P$_B$2$_{1}$2$_{1}$2 and P$_B$2$_{1}$2$_{1}$2. The spin arrangements associated with these groups are shown in Fig.~\ref{fig:state} using a conventional hexagonal chemical unit cell and cell doubling along the a-axis. These models are in one-to-one correspondence to irreducible representations (IRs) of the little group G$_{\bf{k}}$, and are labelled as $\Gamma_{1}$, $\Gamma_{2}$, $\Gamma_{3}$, and $\Gamma_{4}$, respectively. Two of the models, $\Gamma_{1}$ and $\Gamma_{4}$, have collinear spins lying purely along the a-axis. The other two, $\Gamma_{2}$ and $\Gamma_{3}$, allow tilting of spins along the c-axis and form more complex non-collinear arrangements in the bc-plane.
\newline

\clearpage
\section{Refinements of neutron diffraction data}

\begin{figure*}[h]
\includegraphics[width=\textwidth]{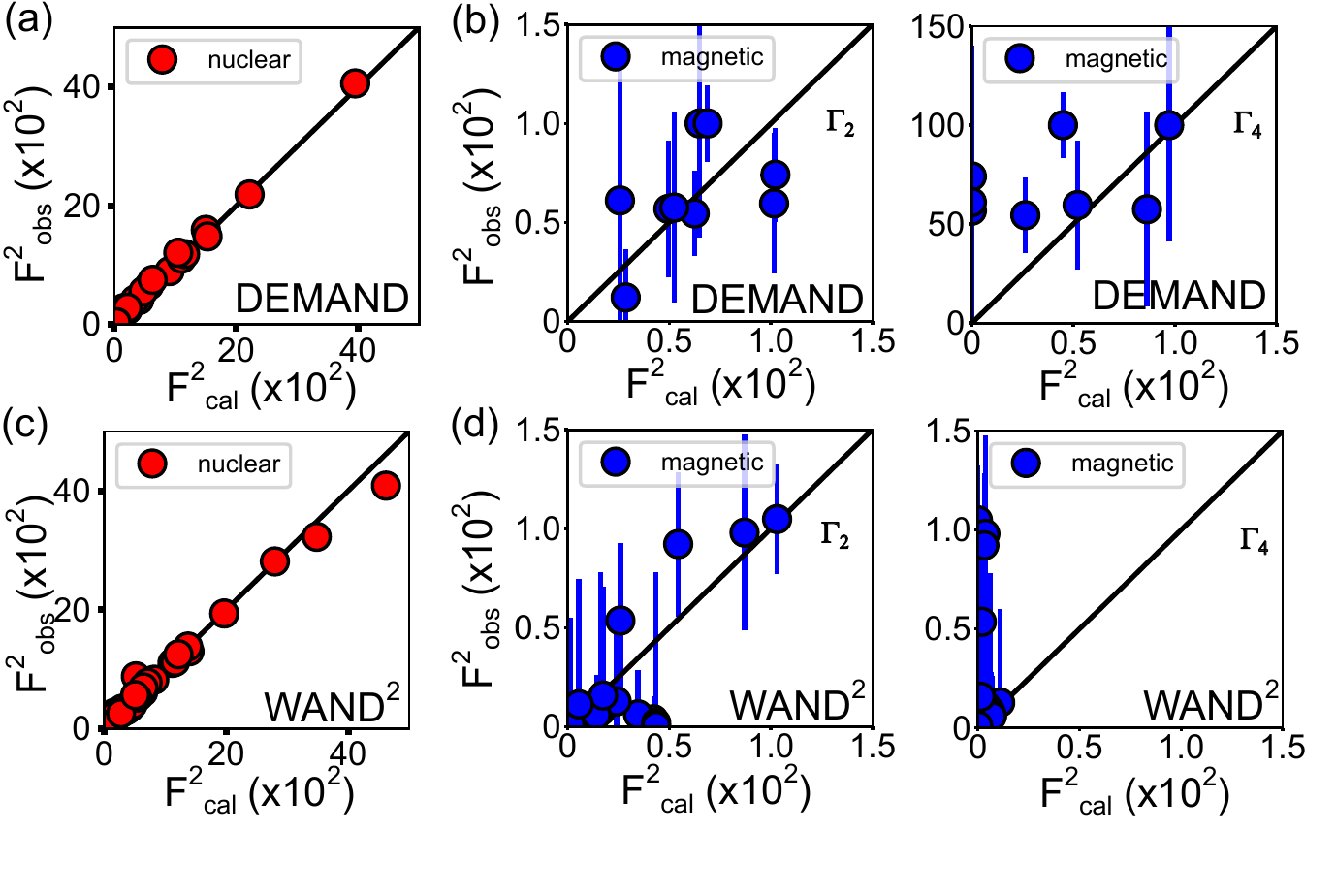}
\caption{\label{fig:fitting} Plots of observed versus calculated structure factors for magnetic (right) and nuclear (left) refinements of neutron scattering data presented in this Letter. Panels (a) and (b) represent fits of data taken using the DEMAND instrument and panels (c) and (d) represent fits of data taken using the WAND$^2$ instrument, both of the HFIR at ORNL. }
\end{figure*}

As described in the main text, diffraction patterns were acquired from our single crystal sample from both DEMAND and WAND$^2$ instruments at the High Flux Isotope Reactor (HFIR) at Oak Ridge National Laboratory. Data from the two instruments were analyzed separately to independently confirm conclusions about the symmetry of the lattice and ordered spin state.  Refinements of the lattice structure were performed assuming P6$_{3}$22 structural model with Co at the 2c Wyckoff sites. Refined sulfur positions were (0.3223, -0.0026, 0.1333) when considering WAND$^2$, and (0.33815753, -0.0972, 0.1329) when considering DEMAND data. Refinements of magnetic structure were performed by considering each of the irreducible representations for the  $\mathbf{k} = (0.5, 0, 0)$ propagation vector in turn, assuming spins on the cobalt sublattice of the lattice structure. For both instruments, fits employing the $\Gamma_2$ model provided the best description of the data. For for data taken with DEMAND, the determination of magnetic Bragg peak intensity was complicated by scattering at the same angular positions from nuclear Bragg peaks with a small number of neutrons with $\lambda/2$. The improvement of $\Gamma_2$ over competing models was marginal due to large statistical error bars. When considering the WAND$^2$ data, the correctness of the $\Gamma_2$ was more definitive. Comparison of calculated and measured Bragg peak intensities for both lattice and magnetic scattering on the two instruments is plotted in Fig.~\ref{fig:fitting}.

\clearpage
\section{Density Functional Theory calculations}

As described in the main text, we used density functional theory (DFT) to check the stability of the magnetic ground state observed with neutron diffraction and calculate the associated Kerr signal. Beginning with the lattice and spin configurations obtained from the neutron analysis, we allowed the system to relax to the lowest energy state. The resultant magnetic order was very similar to that determined by our neutron scattering analysis. The main difference were a slightly smaller ordered moment predicted by DFT and a slight larger tilting angle. The exact DFT predictions for the four moments in the magnetic unit cell are shown in Table \ref{tab:dft}, alongside those determined by neutron diffraction.

\begin{table}[h]
\caption{Non-collinear magnetic ordering of four Co magnetic sites determined from neutron scattering measurement and DFT relaxed result using the experimental values. Magnetization is in Cartesian coordinate with unit $\mu_B$.
}
\begin{center}
\begin{tabular}{c|c c c c}
\hline
Exp. & $M_x$ & $M_y$ & $M_z$ & $|\textbf{M}|$ \\
\hline
Co (\MakeUppercase{\romannumeral 1})& $-0.965$ & 1.671 & 0.358 & 1.962\\
Co (\MakeUppercase{\romannumeral 2})& 0.965 & $-1.671$ & 0.358 &  1.962\\
Co (\MakeUppercase{\romannumeral 3})& $-0.965$ & 1.671 & $-0.358$ &  1.962\\
Co (\MakeUppercase{\romannumeral 4})& 0.965 & $-1.671$ & $-0.358$ & 1.962\\
\hline
DFT & $M_x$ & $M_y$ & $M_z$ & $|\textbf{M}|$ \\
\hline
Co (\MakeUppercase{\romannumeral 1})& $-0.712$ & 1.252 & $-0.103$ & 1.444\\
Co (\MakeUppercase{\romannumeral 2})& 0.723 & $-1.247$ & $-0.102$ & 1.445\\
Co (\MakeUppercase{\romannumeral 3})& $-0.726$ & 1.245 & 0.080 & 1.443\\
Co (\MakeUppercase{\romannumeral 4})& 0.716 & $-1.249$ & 0.124 & 1.445\\
\hline
\end{tabular}
\end{center}
\label{tab:dft}
\end{table}

\begin{figure}
\includegraphics[width=12cm]{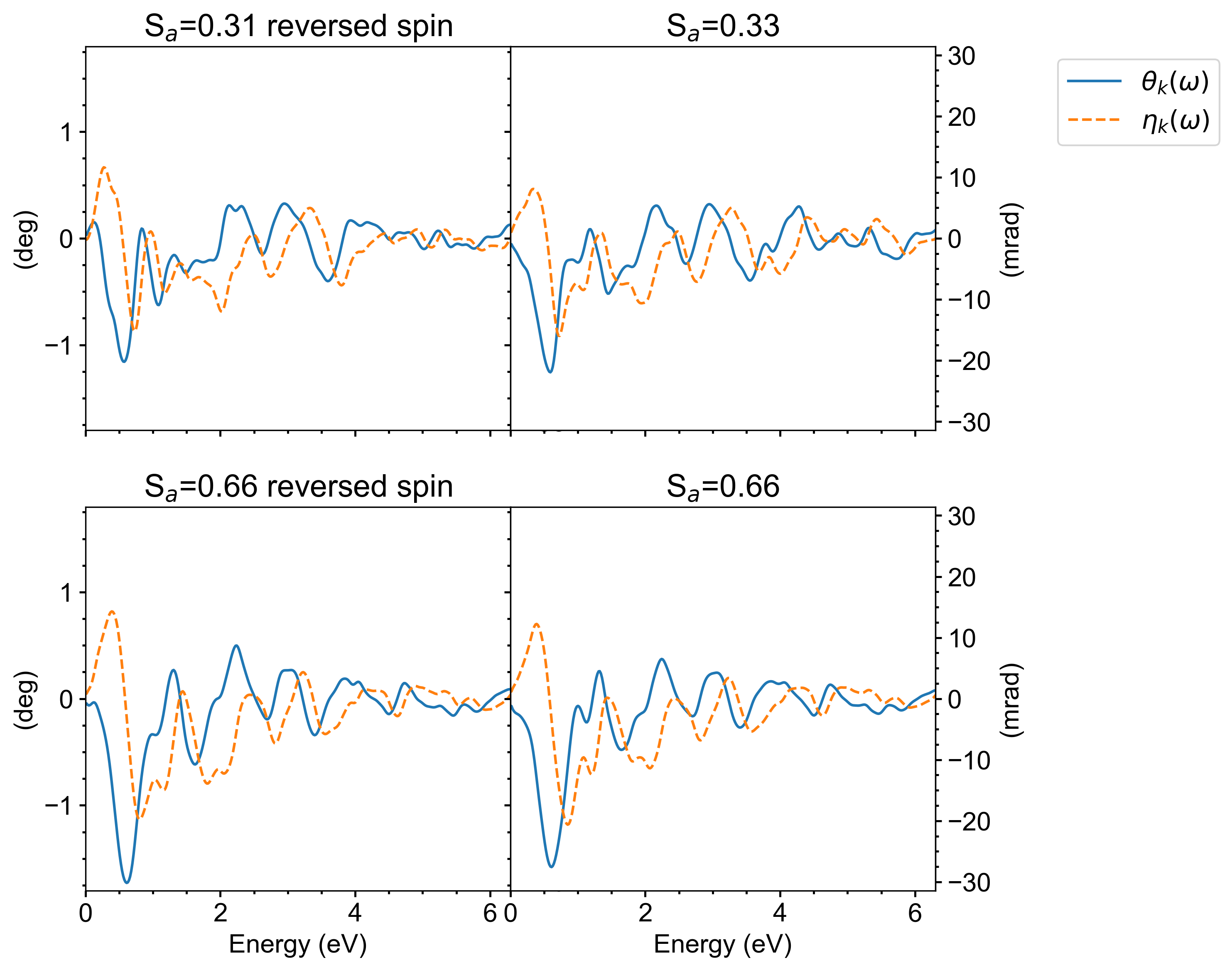}
\caption{\label{fig:fig1} Kerr rotation $\theta$ and ellipticity $\eta$ calculated from density functional theory using four magnetic unit cells based on the refinement of neutron diffraction measurements.
}
\end{figure}

Using the same numerical convergence parameters, we further calculated the Kerr signals for several additional magnetic and lattice configurations for comparison:
Two cases used the same magnetic ordering as discussed in the paper but instead of 0.31 as fractional sulfur coordinate used in the main text, sulfur was located at fractional coordinates of 0.33 and 0.66 of the $a$-lattice dimension, which are the positions for completely undistorted CoS$_6$ octahedra. The two values here represent opposite choices of chirality. For sulfur at 0.31 and 0.66, we generated two more cases by reversing the sign of all magnetic moments.  All four of these additional cases show large Kerr signals of the same order of magnitude as the case presented in the main text. See Fig.\ \ref{fig:fig1} for these results.

\begin{figure}
\includegraphics[width=12cm]{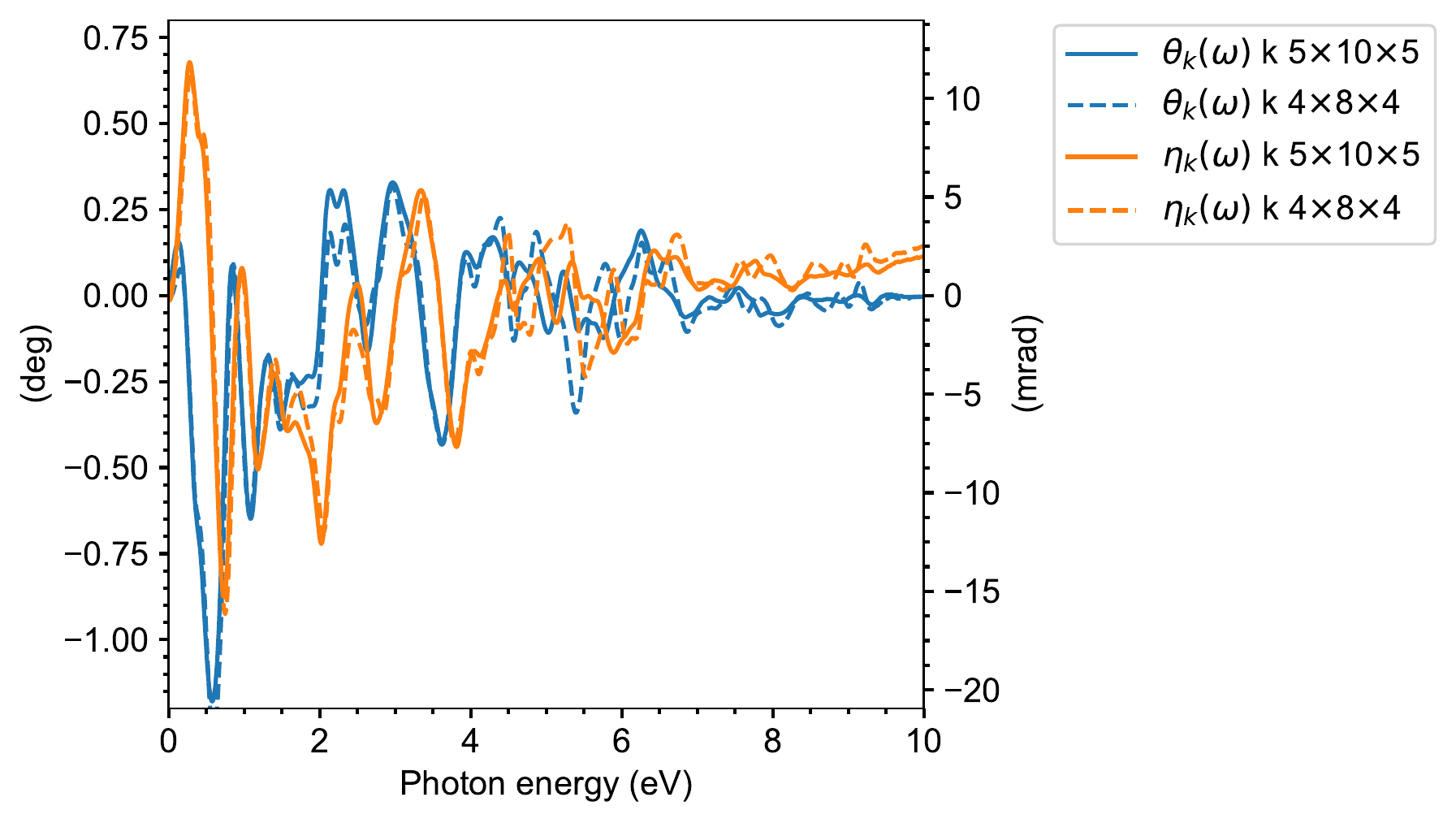}\caption{\label{fig:fig2} $\bf{k}$ convergence of Kerr rotation $\theta$ and ellipticity $\eta$ is shown with 5$\times$10$\times$5 and 4$\times$8$\times$4 Monkhorst-Pack grid.
From this figure we determine the remaining error due to \textbf{k}-point convergence as described in the main text.
}
\end{figure}

In Fig.\ \ref{fig:fig2} we show the \textbf{k}-point convergence of the Kerr signals computed in this work for the atomic geometry and magnetic configuration used in the main text.

\clearpage


%

%

\end{document}